\documentclass[reprint,
superscriptaddress,
groupedaddress,
showpacs,preprintnumbers,
nobibnotes,
amsmath,amssymb,
aps,showkey
pra,
floatfix,
]{revtex4-2}

\usepackage{graphicx}
\usepackage{dcolumn}
\usepackage{bm}
\usepackage{epstopdf}
\usepackage{float}
\usepackage[caption = false]{subfig}
\usepackage{array}
\usepackage{braket}
\usepackage{physics}
\usepackage{graphicx}
\usepackage{cancel}
\usepackage{ulem}
\usepackage{xcolor}
\begin{document}

\title{Nonadiabatic Dynamics and Rotational Coupling in $\mathrm{HeH^+}$ Dissociative Recombination and Resonant Ion-Pair Formation}

\author{Sifiso M. Nkambule}
\email{snkambule@uneswa.ac.sz}
\affiliation{Department of Physics, University of Eswatini, Kwaluseni, M201, Eswatini.} 
\author{Malibongwe Tsabedze}
\affiliation{Department of Physics, University of Eswatini, Kwaluseni, M201, Eswatini.} 
\author{Oscar N. Mabuza}%
\affiliation{Department of Physics, William Pitcher College , Manzini, M200, Eswatini.}%
\affiliation{Department of Physics, University of Eswatini, Kwaluseni, M201, Eswatini.} 
\author{Mbuso K. Matfunjwa}%
%
\affiliation{Department of Physics and Astronomy, University of Nebraska, Lincoln, Nebraska 68588-0299, USA}


\affiliation{Department of Physics, University of Eswatini, Kwaluseni, M201, Eswatini.}

\date{\today}

\begin{abstract}
We present a time-dependent wave-packet study of dissociative recombination (DR) and resonant ion-pair (RIP) formation in $\mathrm{HeH^+}$ isotopologues. Nuclear dynamics are treated on a manifold of 23 coupled electronic states of $^2\Sigma$, $^2\Pi$, and $^2\Delta$ symmetries, including rotational couplings between different symmetries.

The results reveal that inclusion of a large manifold of resonant states and rotational couplings significantly enhances the DR cross section relative to earlier theoretical studies. In the diabatic representation, $^2\Sigma$ states dominate the recombination dynamics, while in the adiabatic representation, $^2\Pi$ and $^2\Delta$ states contribute significantly at low collision energies. For RIP formation, two different diabatization schemes yield systematically larger cross sections than previous models, highlighting the sensitivity of ion-pair production to electronic coupling structure. Isotopic effects are examined, showing a clear inverse dependence of cross section magnitude on reduced mass. Thermal rate coefficients are computed over $10^{2}$ to $2\times 10^4$ K thermal electron temperatures. 

Isotopic effects are examined, showing a clear inverse dependence of cross section magnitude on reduced mass. The results are compared with rotational-state-resolved experimental and theoretical results. 

The present results highlight the importance of multistate coupling and rotational interactions in electron-driven fragmentation processes relevant to primordial and astrophysical plasmas

\end{abstract}

\maketitle


\section{Introduction}
\label{intro}

Dissociative recombination (DR) and resonant ion-pair (RIP) formation are among the most important electron-driven processes governing the evolution of molecular plasmas. Through these reactions, molecular ions are neutralized by low-energy electrons, thereby influencing ionization balance, molecular abundances, energy transfer, and chemical evolution in laboratory, atmospheric, and astrophysical environments~\cite{GalliPalla1998,Lepp02,Tennyson2010,Schneider2017}. Accurate descriptions of these processes are therefore essential for plasma kinetic modelling and for understanding the chemistry of low-temperature ionized gases.

Among molecular ions of astrophysical interest, the helium hydride ion, $\mathrm{HeH^+}$, occupies a unique position. It is widely regarded as the first molecular ion formed in the Universe, originating shortly after cosmological recombination through radiative association reactions involving helium and hydrogen nuclei~\cite{GalliPalla1998,Lepp02,Black76,Stromholm}. Because of its large dipole moment and efficient radiative cooling properties, $\mathrm{HeH^+}$ has long been recognized as a potentially important species in primordial gas chemistry and early structure formation~\cite{GalliPalla1998,Lepp02}. Interest in this molecular ion increased substantially following its astronomical detection in the planetary nebula NGC 7027 by Güsten \textit{et al.}~\cite{Guesten19}, providing direct observational confirmation of a species that had been predicted theoretically for several decades.

The abundance and survival of $\mathrm{HeH^+}$ in astrophysical environments are strongly controlled by electron-impact destruction processes, particularly dissociative recombination. In the DR process,
\begin{equation}
\mathrm{HeH}^{+}+e^{-}\rightarrow \mathrm{He}+\mathrm{H},
\end{equation}

an incident electron is captured into a neutral resonant state of the molecule, followed by dissociation into neutral fragments. Depending on the electronic structure of the system, DR may proceed through either direct or indirect mechanisms~\cite{Bates,GiustiSuzor1980,Guberman}. In the direct mechanism, the electron is captured into a dissociative resonant state leading to prompt fragmentation. In the indirect mechanism, the electron is first captured into a Rydberg state, which subsequently predissociates through coupling to dissociative channels~\cite{Douguet,Guberman,GiustiSuzor1980}. The relative importance of these pathways is highly sensitive to the electronic-state manifold and the nonadiabatic couplings that connect them.

Closely related to dissociative recombination is resonant ion-pair formation,
\begin{equation}
\mathrm{HeH}^{+}+e^{-}\rightarrow \mathrm{He}^{+}+\mathrm{H}^{-},
\end{equation}

which occurs through population transfer from neutral resonant states into ion-pair channels. Although generally less efficient than DR, ion-pair formation can become important in plasmas where negative ions contribute significantly to the chemistry and charge balance~\cite{Lepadellec2001,Zong1999,Larson98,Larson15}. Because the process is mediated by avoided crossings and electronic-state couplings, it provides a sensitive probe of nonadiabatic molecular dynamics~\cite{Mead1982}.

Theoretical investigations of DR in $\mathrm{HeH^+}$ have a long history. Early studies based on multichannel quantum defect theory (MQDT) and R-matrix methodologies established the importance of resonant electron capture and nonadiabatic interactions in determining the recombination probability~\cite{Sarpal94,Guberman,Sarpal1993}. Guberman~\cite{Guberman} demonstrated that indirect recombination mechanisms could significantly enhance the DR rate, while calculations by Sarpal, Tennyson, and Morgan~\cite{Sarpal1993,Sarpal94} provided some of the first quantitative cross sections for the system. Subsequently, Orel, Kulander, and Rescigno~\cite{Orel95} introduced a time-dependent wave-packet treatment that enabled a direct description of nuclear dynamics on transient neutral states. These studies established the wave-packet approach as a powerful tool for investigating electron-molecule recombination processes.

Experimental measurements have likewise revealed substantial recombination rates for $\mathrm{HeH^+}$, despite the relatively simple electronic structure of the ion. Storage-ring experiments performed by Yousif and Mitchell~\cite{Yousif89}, Mowat \textit{et al.}~\cite{Mowat95}, Sundstr\"om \textit{et al.}\cite{Sundstrom94}, and Semaniak \textit{et al.}~\cite{Seminiak96} demonstrated that the DR rate is significantly larger than might be expected from a simple direct-capture picture. These findings highlighted the importance of electronic-state coupling and resonance dynamics in the recombination process.

Larson \textit{et al.}~\cite{Larson98} performed a detailed wave-packet study of both dissociative recombination and ion-pair formation in $\mathrm{HeH^+}$, demonstrating the importance of nonadiabatic interactions near avoided crossings. However, the calculations were restricted to a relatively small set of resonant states and did not include rotational couplings between electronic symmetries. Subsequent electronic-structure calculations by Larson \textit{et al.}~\cite{Larson15,Larson16} revealed a considerably richer manifold of resonant states and a complex network of electronic and rotational couplings involving $^{2}\Sigma$, $^{2}\Pi$, and $^{2}\Delta$ symmetries. These findings suggest that earlier dynamical models may not have fully captured the available dissociation pathways.

The role of rotational excitation has also attracted increasing attention. Recent cryogenic storage-ring measurements and theoretical analyses by Novotný \textit{et al.}~\cite{Novotny2019} demonstrated that rotational-state populations significantly influence the recombination rate of $\mathrm{HeH^+}$. Their results showed systematic variations between rotational levels $j=0$, $j=1$, $j=2$, and $j\ge3$, highlighting the importance of rotational dynamics in determining the overall reaction probability. Nevertheless, a fully coupled treatment incorporating both a large manifold of resonant states and rotational interactions between different electronic symmetries remains lacking.

From a theoretical perspective, the $\mathrm{HeH}$ system represents an ideal test case for studying the interplay between electronic structure, nonadiabatic coupling, rotational interactions, and nuclear motion. The existence of numerous avoided crossings between covalent and ion-pair states, together with strong electronic couplings and significant isotope effects, provides a stringent benchmark for modern wave-packet methodologies~\cite{Hedberg14,Larson98,Larson16,Nkambule22}.

In the present work, we perform a comprehensive investigation of DR and RIP formation in $\mathrm{HeH^+}$ isotopologues using time-dependent wave-packet propagation. Nuclear dynamics are treated on a manifold of 23 coupled resonant electronic states of $^{2}\Sigma$, $^{2}\Pi$, and $^{2}\Delta$ symmetries. Rotational couplings between states of different symmetries are included explicitly, and calculations are carried out in both adiabatic and strictly diabatic representations~\cite{Mead1982}. This approach extends previous treatments by incorporating a significantly larger electronic-state manifold together with rotationally induced population transfer mechanisms.

The objectives of this work are threefold. First, we assess the influence of rotational couplings and extended electronic-state interactions on dissociative recombination cross sections. Second, we investigate the sensitivity of resonant ion-pair formation to different diabatization schemes and electronic-coupling structures. Third, we examine isotope effects and derive thermal rate coefficients relevant to laboratory plasmas, planetary nebulae, diffuse interstellar clouds, and primordial gas chemistry. By comparing our results with previous theoretical studies~\cite{Sarpal94,Larson98} and experimental measurements~\cite{Mowat95,Yousif89,Novotny2019}, we provide new insight into the mechanisms governing electron-driven fragmentation of $\mathrm{HeH^+}$ and its isotopologues.

The remainder of this paper is organized as follows. Section~\ref{pec} describes the electronic structure calculations, diabatic representations, and rotational couplings. Section~\ref{nd} outlines the time-dependent wave-packet methodology used to compute cross sections. Section~\ref{dresults} presents and discusses the calculated dissociative recombination and ion-pair formation cross sections, isotope effects, and symmetry-resolved contributions. Thermal rate coefficients and their astrophysical implications are presented in Section~\ref{trates}. Finally, conclusions are given in Section~\ref{concl}. 

\section{\label{pec}POTENTIAL ENERGY CURVES AND ELECTRONIC COUPLINGS}

\subsection{Electronic Structure and Resonant States}

The present study employs potential energy curves and electronic coupling elements obtained from high-level \textit{ab initio} electronic structure calculations combined with electron scattering methods. The potential curves used in this work were first reported by Larson \textit{et al.}~\cite{Larson15,Larson16}, where the electronic structure was computed at the full configuration interaction (FCI) level of theory. This approach provides an accurate description of both the bound and resonant electronic states of the neutral HeH system, which is essential for describing the dissociative recombination dynamics.

The FCI calculations were performed over a range of internuclear distances, \(R\), relevant to the dissociation dynamics. The resulting potential energy curves describe the adiabatic states of the neutral molecule, including the resonant states that lie energetically above the ground state of the $\mathrm{HeH+}$ ion. These resonant states are the key intermediates in the dissociative recombination process, as they provide the temporary capture states for the incident electron~\cite{Orel2000}.

To obtain the autoionization widths \(\Gamma_i(R)\) and the positions of the resonant states relative to the ionic ground state, electron scattering calculations were performed using the complex-Kohn variational method~\cite{Rescigno1994, Rescigno1995}. This approach is particularly well-suited for treating electron-molecule scattering in the resonance region because it provides a rigorous treatment of the scattering wavefunction and the associated resonance parameters. The complex-Kohn method yields both the energy positions of the resonances and their autoionization widths as functions of the internuclear distance.

The autoionization widths \(\Gamma_i(R)\) are non-zero only at short and intermediate internuclear distances, where the resonant states have significant overlap with the ionization continuum. At large internuclear distances, the resonant states become purely dissociative and \(\Gamma_i(R) \to 0\). This distance dependence is physically important because it determines the competition between dissociation and autoionization: the wave packet must reach the region where \(\Gamma_i(R)\) becomes small to avoid re-ionization and successfully dissociate into neutral fragments.

For clarity, the adiabatic potential energy curves and autoionization widths are shown in figures 1 and 4(a) of Ref.~\cite{Larson15}, respectively. The ground-state potential energy curve of the $\mathrm{HeH+}$ ion, which provides the initial vibrational wavefunction for the wave-packet propagation, is also included in these calculations.

\subsection{Adiabatic and Diabatic Representations}

In the adiabatic representation, the electronic wavefunctions are eigenfunctions of the electronic Hamiltonian at each internuclear distance \(R\). Consequently, the potential energy curves are smooth functions of \(R\), but the electronic character of the states changes abruptly at avoided crossings. The adiabatic states of \(^2\Sigma\) symmetry in the HeH system do not preserve their character across the full range of internuclear distances~\cite{Larson16, Larson15}.

Specifically, the lowest-energy \(^2\Sigma\) state has ion-pair character ($\mathrm{He^+ + H^-}$) at short internuclear distances, while at large internuclear distances, this ion-pair character is exhibited by the highest-energy \(^2\Sigma\) state. Between these limits, the states undergo a series of avoided crossings where the ion-pair and covalent characters are exchanged. Whenever there is a switch from an ion-pair state to a covalent state, avoided crossings are observed~\cite{Larson16}. These avoided crossings are the primary regions where nonadiabatic couplings become large and where population transfer between electronic states can occur.

In contrast, the diabatic representation is defined such that the electronic character of each state is preserved along the entire internuclear coordinate~\cite{Mead1982}. This is achieved by performing an adiabatic-to-diabatic transformation that removes the derivative couplings responsible for the abrupt changes in electronic character. In the diabatic basis, the potential energy curves may cross each other, but the physical interpretation of each state remains clear throughout the dissociation process~\cite{Larson16, Nkambule22}.

The strict adiabatic-to-diabatic transformation of 23 coupled states, following the method of Larson \textit{et al.}~\cite{Larson16}, provides a diabatic representation where:
\begin{enumerate}
\item the ion-pair and covalent characters are preserved on each potential energy curve,
\item the potential energy curves cross each other at the locations of the avoided crossings,
\item the electronic couplings between states of the same symmetry vary smoothly with \(R\).
\end{enumerate}

In the diabatic representation, the lowest resonant state at short internuclear distances crosses the other resonant states and dissociates to the ion-pair limit $\mathrm{He^+ + H^-}$, while the other covalent states dissociate to neutral fragments. This representation is particularly useful for describing the resonant ion-pair (RIP) formation process, where population must be transferred from covalent states into the ion-pair channel through the avoided crossings.

\subsection{Diabatic Representation for RIP Formation}

\begin{figure}[!h]
\includegraphics[scale=0.35,angle=-90]{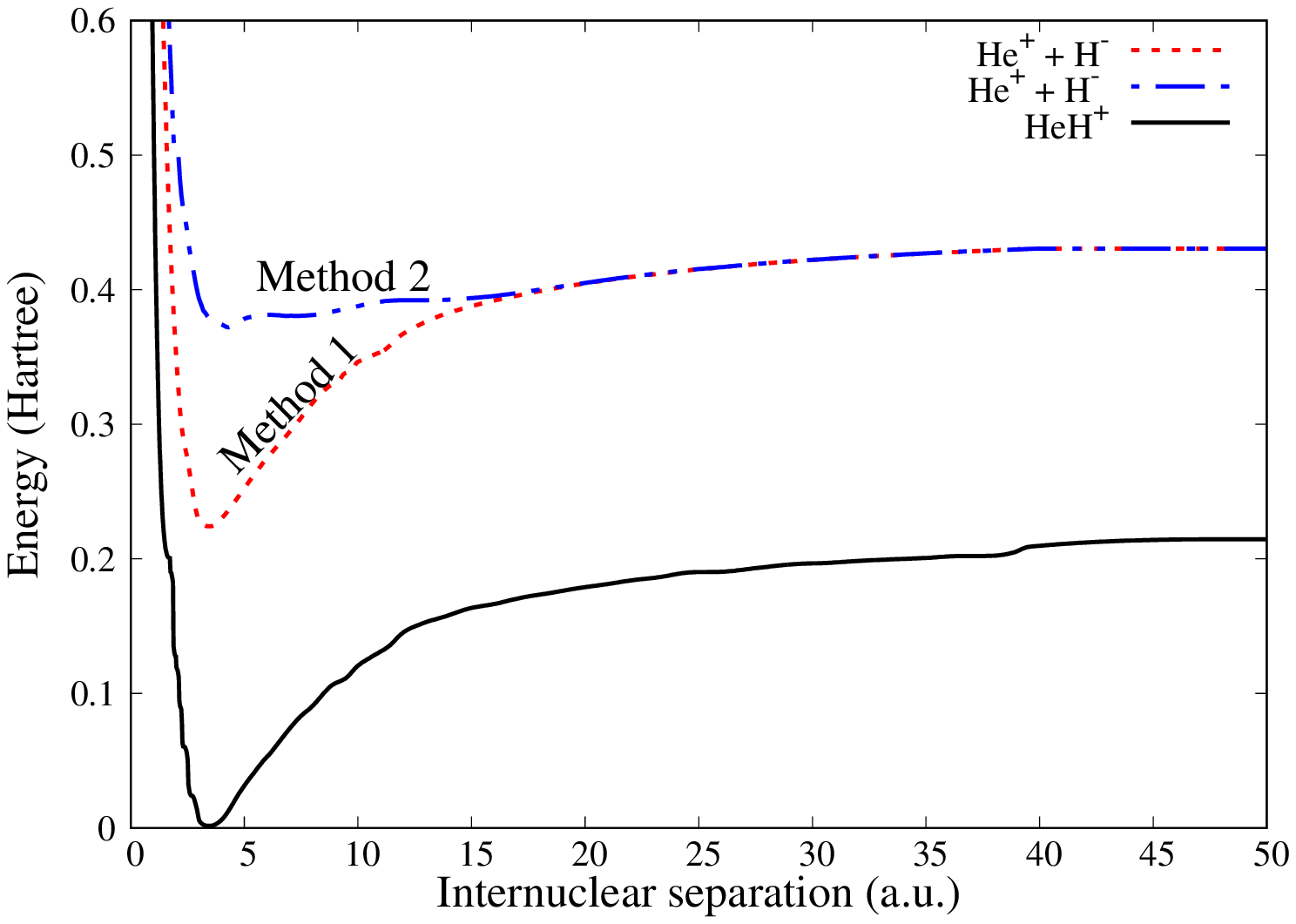}
\caption{\label{ip1} Potential energy curves for the $\mathrm{He^+ + H^-}$ states in the two diabatization representations and the ground state of $\mathrm{HeH^+}$.}
\end{figure}

For the resonant ion-pair formation process, two different diabatization methods are employed in this work, as illustrated in Figure~\ref{ip1}.

\textbf{Method 1: Two-state diabatization.} In this approach, the wave packets are propagated on an ion-pair potential energy curve obtained from a two-by-two adiabatic-to-diabatic transformation~\cite{Hedberg14,Nkambule22}. Only two adiabatic states are assumed to be interacting near a single avoided crossing, and all other states are neglected in the transformation. The ion-pair and covalent characters are preserved for the two interacting states, and the resulting ion-pair potential energy curve is labelled \textbf{Method 1} in Figure~\ref{ip1}.

\textbf{Method 2: Strict 23-state diabatization.} In this approach, the wave packets are propagated on a potential energy curve obtained from the strict adiabatic-to-diabatic transformation of all 23 coupled states~\cite{Larson16}. The transformation is performed simultaneously for all states, preserving the ion-pair and covalent character on each of the potential energy curves. The ion-pair character is associated with the potential energy curve that is highest in energy at large internuclear distances, and this curve is labelled \textbf{Method 2} in Figure~\ref{ip1}.

The comparison between these two methods allows us to assess the importance of multistate coupling effects in the ion-pair formation process. As will be shown in Section IV, the strict 23-state diabatization yields significantly larger ion-pair cross sections than the two-state approximation, indicating that the cumulative effect of multiple avoided crossings plays a crucial role in the population transfer dynamics.

All potential energy curves are shifted such that the minimum energy of the ground state of the $\mathrm{HeH^+}$ ion is set to zero. This convention simplifies the interpretation of the scattering energies and ensures consistent energy referencing across the calculations.

\subsection{Rotational Couplings}

An important extension of the present work is the explicit inclusion of rotational couplings between electronic states of different symmetries. The \(\mathbf{L}\)-uncoupling term of the rotational Hamiltonian, which describes the interaction between the electronic orbital angular momentum and the molecular rotation, can be written as~\cite{LefebvreBrion1986}:

\begin{equation}
\begin{aligned}
&-\frac{1}{2\mu R^2}\langle J,S,\Omega \pm 1,\Lambda,\Sigma \pm 1|J_{\pm}L_{\pm}|J,S,\Omega,\Lambda,\Sigma\rangle \\
&= -\frac{1}{2\mu}\left[J(J+1) - \Omega(\Omega \pm 1)\right]^{1/2} \\
&\qquad \times \langle J,S,\Omega \pm 1,\Lambda,\Sigma \pm 1|L_{\pm}|J,S,\Omega,\Lambda,\Sigma\rangle,
\end{aligned}
\tag{3}
\end{equation}

where \(J\), \(L\), and \(S\) are the total angular momentum, total orbital angular momentum, and total spin quantum numbers, respectively. The quantities \(\Omega\), \(\Lambda\), and \(\Sigma\) are their respective projections onto the molecular axis. The operators \(J_{\pm}\) and \(L_{\pm}\) are the raising and lowering operators for the total and orbital angular momenta.

The molecular electronic wavefunctions for the \(^2\Sigma^+\), \(^2\Pi\), and \(^2\Delta\) states are denoted as \(\Phi_{\Sigma}\), \(\Phi_{\Pi}\), and \(\Phi_{\Delta}\), respectively. Based on the selection rules for the \(\mathbf{L}\)-uncoupling operator, the allowed transitions between symmetries satisfy \(\Delta\Lambda = \pm 1\). Consequently, rotational couplings between \(\Phi_{\Sigma}\) and \(\Phi_{\Delta}\) are zero because \(\Delta\Lambda = 2\) is forbidden.

Neglecting the small term \(\Omega(\Omega+1)\) compared with \(\ell(\ell+1)\), the rotational coupling between \(\Phi_{\Sigma}\) and \(\Phi_{\Pi}\) states simplifies to:

\begin{equation}
-\frac{[\ell(\ell+1)]^{1/2}}{2\mu R^2}\langle \Phi_{\Pi}|L_{\pm}|\Phi_{\Sigma}\rangle,
\tag{4}
\end{equation}

where \(\ell\) are the rotational quantum numbers. This approximation is justified because \(\Omega(\Omega+1)\) is of order 1/4 and is negligible compared with the typical values of \(\ell(\ell+1)\) encountered in the calculations.

From the electronic structure calculations of the HeH system~\cite{Larson15}, the dominant configurations for covalent states associated with the same asymptotic limit have the form \((1\sigma)^1(2\sigma)^1(n\lambda)^1\). Using these configurations, the electronic states of different symmetries are related by:

\begin{equation}
\langle \Phi_{\Pi}|L_{\pm}|\Phi_{\Sigma}\rangle = \langle (np\pi)^1|\ell_+|(np\sigma)^1\rangle,
\tag{6a}
\end{equation}

\begin{equation}
\langle \Phi_{\Delta}|L_{\pm}|\Phi_{\Pi}\rangle = \langle (np\delta)^1|\ell_+|(np\pi)^1\rangle,
\tag{6b}
\end{equation}

where \(\ell_+\) is the one-electron raising operator obtained from the decomposition of \(L_+\).

Applying the pure precession approximation~\cite{vanHemert1991}, which assumes that the electronic orbital angular momentum is dominated by a single molecular orbital, Eqs.~(6a) and (6b) reduce to:

\begin{equation}
\langle \Phi_{\Pi}|L_{\pm}|\Phi_{\Sigma}\rangle = \sqrt{2},
\tag{7}
\end{equation}

\begin{equation}
\langle \Phi_{\Delta}|L_{\pm}|\Phi_{\Pi}\rangle = \sqrt{2}.
\tag{8}
\end{equation}

For covalent states associated with different asymptotic limits, the rotational couplings vanish:

\begin{equation}
\langle \Phi_{\Pi}|L_{\pm}|\Phi_{\Sigma}\rangle = 0,
\tag{10a}
\end{equation}

\begin{equation}
\langle \Phi_{\Delta}|L_{\pm}|\Phi_{\Pi}\rangle = 0.
\tag{10b}
\end{equation}

Based on the above discussion, the adiabatic potential energy curves are represented by the electronic potential energies obtained from the FCI calculations~\cite{Larson15}. In addition to the diagonal potential energies, rotational couplings introduce off-diagonal matrix elements of the form:

\begin{equation}
V_{\Pi,\Sigma}^{a,\ell}(R) = -\frac{\sqrt{\ell(\ell+1)}}{2\mu R^2},
\tag{11}
\end{equation}

which depend on the rotational quantum number \(\ell\) and the internuclear distance \(R\). These couplings connect states of different symmetries and provide an additional mechanism for population transfer that was neglected in earlier studies.

The strict adiabatic-to-diabatic transformation is carried out with the inclusion of rotational couplings to obtain diabatic potential energy curves and electronic couplings, following the procedure described by Larson \textit{et al.}~\cite{Larson16}. This approach ensures that both the nonadiabatic couplings arising from avoided crossings and the rotational couplings between different symmetries are treated consistently in both representations.

\section{\label{nd}NUCLEAR DYNAMICS}

\subsection{Time-Dependent Wave-Packet Method}

The nuclear dynamics of the dissociative recombination and ion-pair formation processes are studied by solving the time-dependent Schr\"odinger equation for the nuclear wave packet:

\begin{equation}
i\frac{\partial}{\partial t}\Psi(R,t) = H_T \Psi(R,t),
\tag{12}
\end{equation}

where \(H_T\) is the total Hamiltonian describing the nuclear motion on the coupled resonant potential energy surfaces. Equation~(12) is solved by direct integration using a wave-packet propagation scheme with the local complex potential~\cite{Birtwistle71, Birtwistle1971b}. This approach treats the autoionization of the resonant states as an imaginary potential that removes probability from the resonant channel and accounts for the competition between dissociation and electron loss.

For states \(i\) and \(j\), the Hamiltonian matrix elements are given by:

\begin{equation}
\begin{aligned}
H_{Tij} = &\left(-\frac{1}{2\mu}\frac{\partial^2}{\partial R^2} + V_i(R) - \frac{i}{2}\Gamma_i(R) - \frac{\sqrt{\ell(\ell+1)}}{2\mu R^2}\right)\delta_{ij} \\
&+ H_{ij}(R),
\end{aligned}
\tag{13}
\end{equation}

where:
\begin{itemize}
\item \(\mu\) is the reduced mass of the system,
\item \(V_i(R)\) is the potential energy of the \(i\)-th resonant state,
\item \(\Gamma_i(R)\) is the autoionization width of the \(i\)-th resonant state,
\item \(\ell\) is the rotational quantum number,
\item the term \(\frac{\sqrt{\ell(\ell+1)}}{2\mu R^2}\) represents the centrifugal barrier associated with the rotational motion,
\item \(H_{ij}(R)\) describes the couplings between states \(i\) and \(j\).
\end{itemize}

The potential energy \(V_i(R)\) can be used in either the adiabatic or diabatic representation, depending on the calculation. In the adiabatic representation, the off-diagonal coupling \(H_{ij}(R)\) corresponds to the second-order derivative nonadiabatic coupling:

\begin{equation}
H_{ij}(R) = -\frac{1}{2\mu}\langle \Phi_i^a(r,R)|\frac{\partial^2}{\partial R^2}|\Phi_j^a(r,R)\rangle,
\tag{14}
\end{equation}

where \(\Phi_i^a(r,R)\) are the adiabatic electronic wavefunctions and \(r\) denotes the electronic coordinates. The nonadiabatic coupling elements for the electronic states of the HeH system are shown in figure 3 of Ref.~\cite{Larson15} and figures 4 and 5 of Ref.~\cite{Larson16}.

In the diabatic representation, by definition~\cite{Mead1982}, the derivative coupling between electronic states vanishes:

\begin{equation}
-\frac{1}{2\mu}\langle \Phi_i^d(r,R)|\frac{\partial^2}{\partial R^2}|\Phi_j^d(r,R)\rangle = 0.
\tag{15}
\end{equation}

In this representation, the coupling \(H_{ij}(R)\) in Eq.~(13) corresponds to the electronic coupling between states of the same symmetry, given by:

\begin{equation}
H_{ij}(R) = \langle \Phi_i^d(r,R)|\mathbf{H}_{el}|\Phi_j^d(r,R)\rangle,
\tag{16}
\end{equation}

where \(\Phi_i^d(r,R)\) are the diabatic electronic wavefunctions and \(\mathbf{H}_{el}\) is the electronic Hamiltonian. Although the adiabatic states \(\Phi_i^a(r,R)\) are eigenfunctions of \(\mathbf{H}_{el}\), the diabatic states \(\Phi_i^d(r,R)\) are not, giving rise to the off-diagonal elements of Eq.~(16). The electronic couplings in the diabatic representation vary smoothly with \(R\) compared with the nonadiabatic couplings of Eq.~(14), making the diabatic representation computationally more stable for multistate calculations~\cite{Hedberg14, Nkambule22, Nkambule2015}.

The autoionization widths \(\Gamma_i(R)\) are obtained from the electron scattering calculations~\cite{Larson15} and are included as a local complex potential within the ``boomerang'' model~\cite{Herzenberg68, Herzenberg79}. This model assumes that the autoionization rate is a local function of the internuclear distance, which is valid when the resonance widths are small compared with the nuclear kinetic energy.

The complex potential approach treats the resonant state as a metastable state with a finite lifetime determined by \(\Gamma_i(R)\). As the wave packet propagates on the resonant potential energy surface, probability is removed from the resonant channel at a rate proportional to \(\Gamma_i(R)\), representing the loss of the electron back to the ionization continuum. The remaining probability that reaches the asymptotic region corresponds to dissociation.

\subsection{Initial Wave Packet and Propagation}

The initial condition for the wave packet is given by~\cite{Orel95}:

\begin{equation}
\Psi(R,t=0) = \sqrt{\frac{\Gamma_i(R)}{2\pi}}\chi_{v=0}(R),
\tag{17}
\end{equation}

where \(\chi_{v=0}(R)\) is the vibrational wavefunction for the \(v=0\) vibrational level of the ground state of $\mathrm{HeH^+}$. This form of the initial wave packet represents the Franck-Condon transition from the vibrational ground state of the ion to the resonant neutral state, with the amplitude weighted by the square root of the autoionization width. The weighting by \(\sqrt{\Gamma_i(R)}\) reflects the fact that electron capture is most probable at internuclear distances where the autoionization width is large.

The vibrational wavefunction \(\chi_{v=0}(R)\) is numerically evaluated using a finite difference method~\cite{Truhlar1972} to solve the time-independent Schr\"odinger equation for the potential energy curve of the ground state of the ion. This approach provides an accurate representation of the initial nuclear wave packet that will be launched onto the resonant potential energy surfaces.

The wave packet is then propagated on the potential energy curves using a numerical algorithm based on the Crank-Nicholson propagation method~\cite{Goldberg1967}. This method is implicit and unconditionally stable, allowing large time steps while maintaining numerical accuracy. The propagation is performed in both the adiabatic and diabatic representations, allowing a direct comparison of the dynamics in the two representations.

\subsection{Cross Section Calculations}

The contributions to the DR reaction and RIP formation cross sections from resonant state \(i\) are computed using:

\begin{equation}
\sigma_i(E) = \frac{2\pi^3}{E}\beta_i |Tr_i(E)|^2,
\tag{18}
\end{equation}

where:
\begin{itemize}
\item \(E\) is the electron scattering energy,
\item \(\beta_i\) is the multiplicity ratio for the final and initial states, accounting for the statistical weights of the electronic states,
\item \(Tr_i(E)\) is the transition amplitude~\cite{Gertitschke1993}.
\end{itemize}

The transition amplitude \(Tr_i(E)\) is obtained by performing a half Fourier transform of the wave packet evaluated at the asymptotic region \(R_{asy}\):

\begin{equation}
Tr_i(E) = \sqrt{\frac{K}{2\pi\mu}}\int_0^{\infty}\Psi_i(R_{asy},t)e^{iEt}dt,
\tag{19}
\end{equation}

where \(K\) is the wave number associated with the dissociating fragments and \(\mu\) is the reduced mass of the system. This Fourier transform projects the time-dependent wave packet onto energy-resolved scattering states, providing the energy-dependent transition amplitude for dissociation.

The factor \(\sqrt{K/(2\pi\mu)}\) in Eq.~(19) ensures the correct normalization of the scattering wavefunction and accounts for the density of states in the continuum. The integral is evaluated numerically using a fast Fourier transform or direct integration, depending on the length of the propagation.

The total DR or RIP cross section is obtained by summing the contributions from all resonant states \(i\):

\begin{equation}
\sigma_{\text{tot}}(E) = \sum_i \sigma_i(E).
\tag{20}
\end{equation}

This summation includes contributions from all symmetries (\(^2\Sigma\), \(^2\Pi\), and \(^2\Delta\)) and all coupled electronic states. The rotational couplings between states of different symmetries are included in the Hamiltonian through the off-diagonal elements of Eq.~(13), ensuring that population transfer between symmetries is properly accounted for.

The present approach extends previous time-dependent wave-packet studies~\cite{Larson98,Orel95} in three important ways:
\begin{enumerate}
\item the inclusion of a manifold of 23 coupled resonant states,
\item the explicit treatment of rotational couplings between states of different electronic symmetries,
\item the use of both adiabatic and strictly diabatic representations for the nuclear dynamics.
\end{enumerate}

These extensions allow a comprehensive investigation of the role of multistate coupling and rotational interactions in the electron-driven fragmentation of $\mathrm{HeH^+}$.

\section{\label{dresults} Results}
\subsection{DR Reaction Cross Section for $\mathrm{^4HeH^+}$}
\begin{figure}[!h]
\includegraphics[scale=0.35,angle=-90]{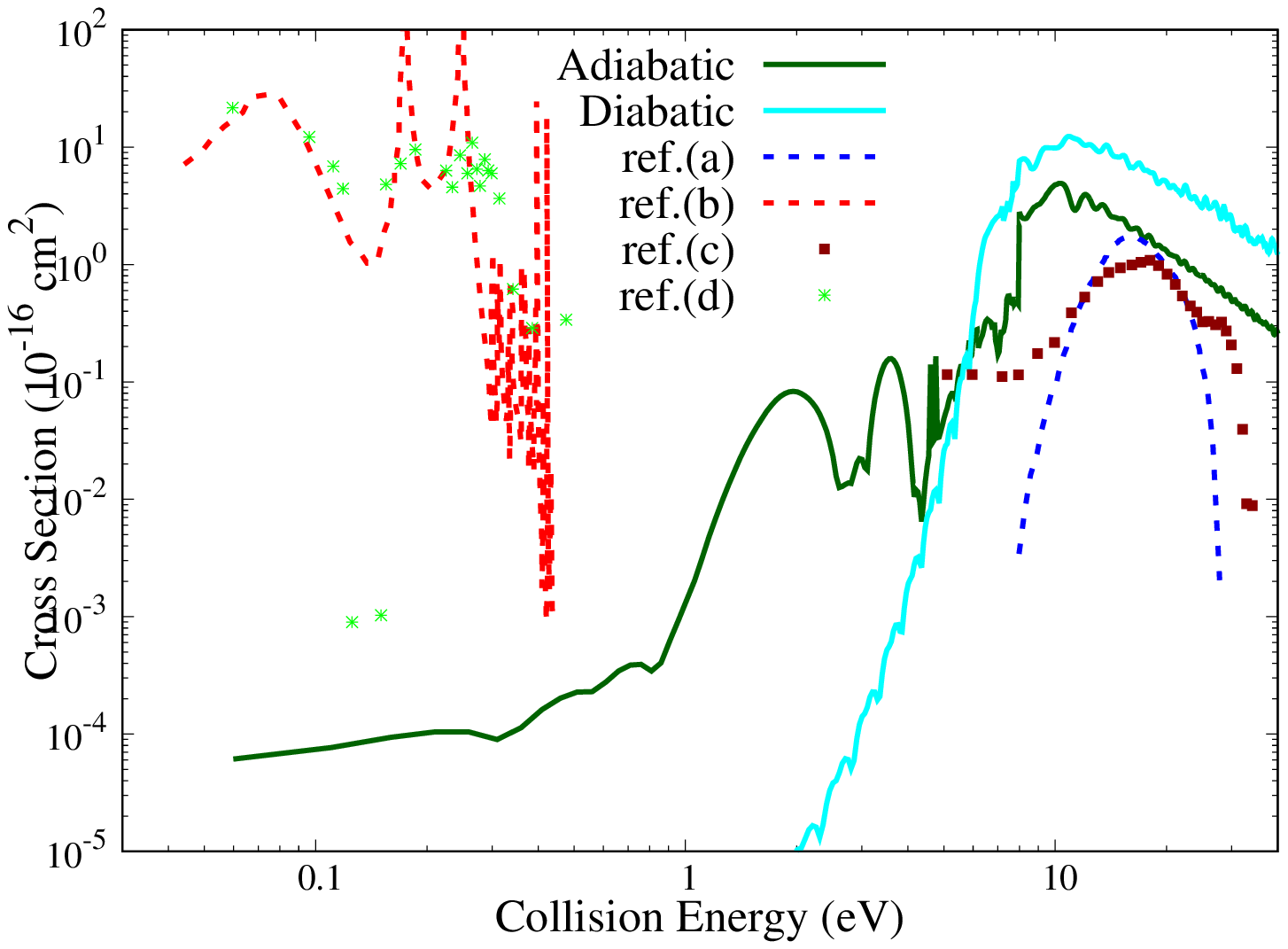}
\caption{\label{cross1} Total DR reaction cross section for $\mathrm{HeH^+}$ as a function of collision energy. Results obtained in both adiabatic and diabatic representation for the different symmetries, in the adiabatic representation. The results are compared with previous theoretical ((a)~\cite{Sarpal94} and (b)~\cite{Larson98}) and Experimental results((c)~\cite{Mowat95} and (d)~\cite{Yousif89})}
\end{figure}

Figure~\ref{cross1} presents the total DR cross section for $\mathrm{^4HeH^+}$ obtained using both adiabatic and diabatic representations and compares the results with previous theoretical and experimental studies. At above 5 eV, the calculated cross sections exhibit the characteristic inverse-energy dependence commonly observed in electron-molecular ion recombination processes, with large values at low collision energies followed by a gradual decrease as the electron energy increases~\cite{Wigner48}.

The present calculations predict substantially larger DR cross sections than earlier wave-packet calculations by Larson \textit{et al.}~\cite{Larson98} and Sarpal \textit{et al.}~\cite{Sarpal94}.

The enhancement originates primarily from the inclusion of a significantly larger manifold of resonant states (23 electronic states compared with only a few states in previous models) and from the explicit incorporation of rotational couplings between states of different electronic symmetries. Similar enhancements arising from additional nonadiabatic pathways have previously been reported in theoretical studies of molecular-ion recombination processes by Orel \textit{et al.}~\cite{Orel95} and Schneider \textit{et al.}~\cite{Schneider2017}.

Several narrow structures and oscillatory features are visible throughout the energy range. These structures are signatures of temporary electron trapping within metastable neutral resonant states before dissociation occurs. Such resonance structures have been identified previously in $\mathrm{HeH^+}$ and other molecular ions and are associated with interference between competing capture and autoionization pathways~\cite{Guberman,Larson98,Motapon06}. The oscillatory behaviour observed above approximately 10 eV is characteristic of energy-dependent electron capture probabilities and has been reported in earlier time-dependent wave-packet studies~\cite{Roos08}.

A notable difference between the two representations is the energy dependence of the cross section magnitude. In the adiabatic representation, the calculated cross sections exhibit significant contributions beginning at lower collision energies, with a gradual rise to a broad maximum before decreasing at higher energies. In contrast, the diabatic representation yields cross sections that are systematically smaller at low energies and reach their maximum at higher collision energies. This behavior reflects the different treatment of avoided crossings and nonadiabatic interactions in the two representations.
In the adiabatic picture, the nonadiabatic couplings are embedded in the nuclear derivative couplings [Eq.~(14)], which efficiently couple the resonant states at the locations of avoided crossings. This allows population transfer to occur over a broader range of internuclear distances, facilitating dissociation even at relatively low collision energies. The derivative couplings are sharply peaked at the avoided crossings and can induce significant transitions even when the nuclear kinetic energy is modest, thereby enhancing the low-energy recombination probability.

n the diabatic representation, the effects of avoided crossings are localized into explicit electronic couplings between states of the same symmetry [Eq.~(16)]. These couplings vary smoothly with internuclear distance and are generally weaker than the sharply peaked derivative couplings in the adiabatic representation. Consequently, population transfer between resonant states is less efficient at low collision energies, requiring higher nuclear kinetic energies to overcome the effective barriers introduced by the diabatic coupling topology. This results in a shift of the dissociative flux to higher collision energies and a corresponding reduction in the low-energy cross section.

The magnitude difference between the two representations is therefore a direct consequence of how the nonadiabatic interactions are represented. The adiabatic representation concentrates the coupling strength into narrow regions around avoided crossings, while the diabatic representation distributes the coupling over a wider range of internuclear distances. The integrated effect on the dissociation probability is sensitive to the energy-dependent dynamics of the nuclear wave packet, leading to the observed differences in the energy dependence of the cross sections.

Most importantly, the present calculations show better agreement with the upper experimental limits reported by Mowat \textit{et al.}~\cite{Mowat95} and Yousif and Mitchell~\cite{Yousif89} than previous theoretical models. This suggests that multi-state coupling effects play a more important role in the recombination dynamics of $\mathrm{HeH^+}$ than previously assumed.
\subsection{Symmetry Contributions to the DR Process}
\begin{figure}[!h]
\includegraphics[scale=0.35,angle=-90]{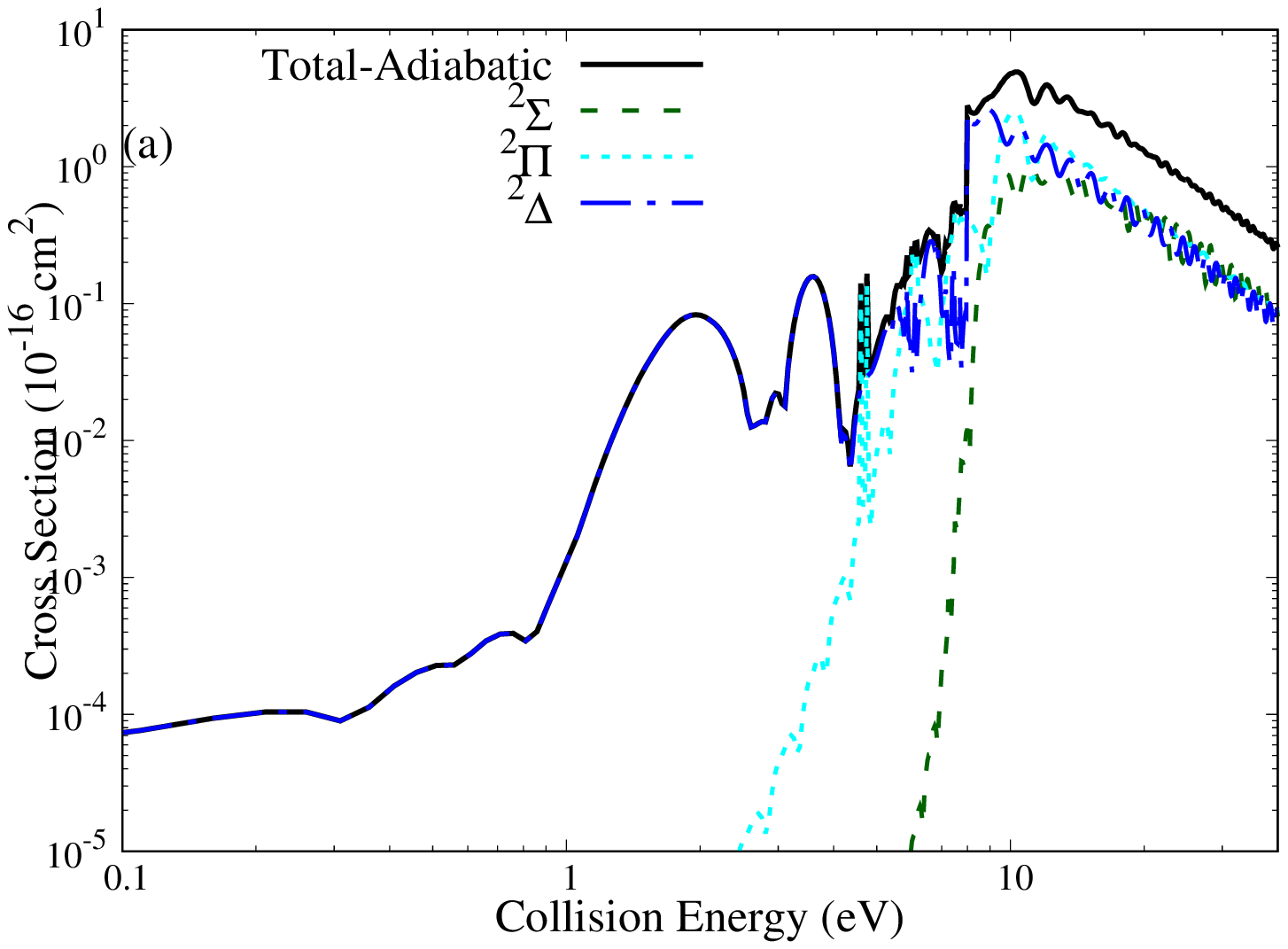}
\includegraphics[scale=0.35,angle=-90]{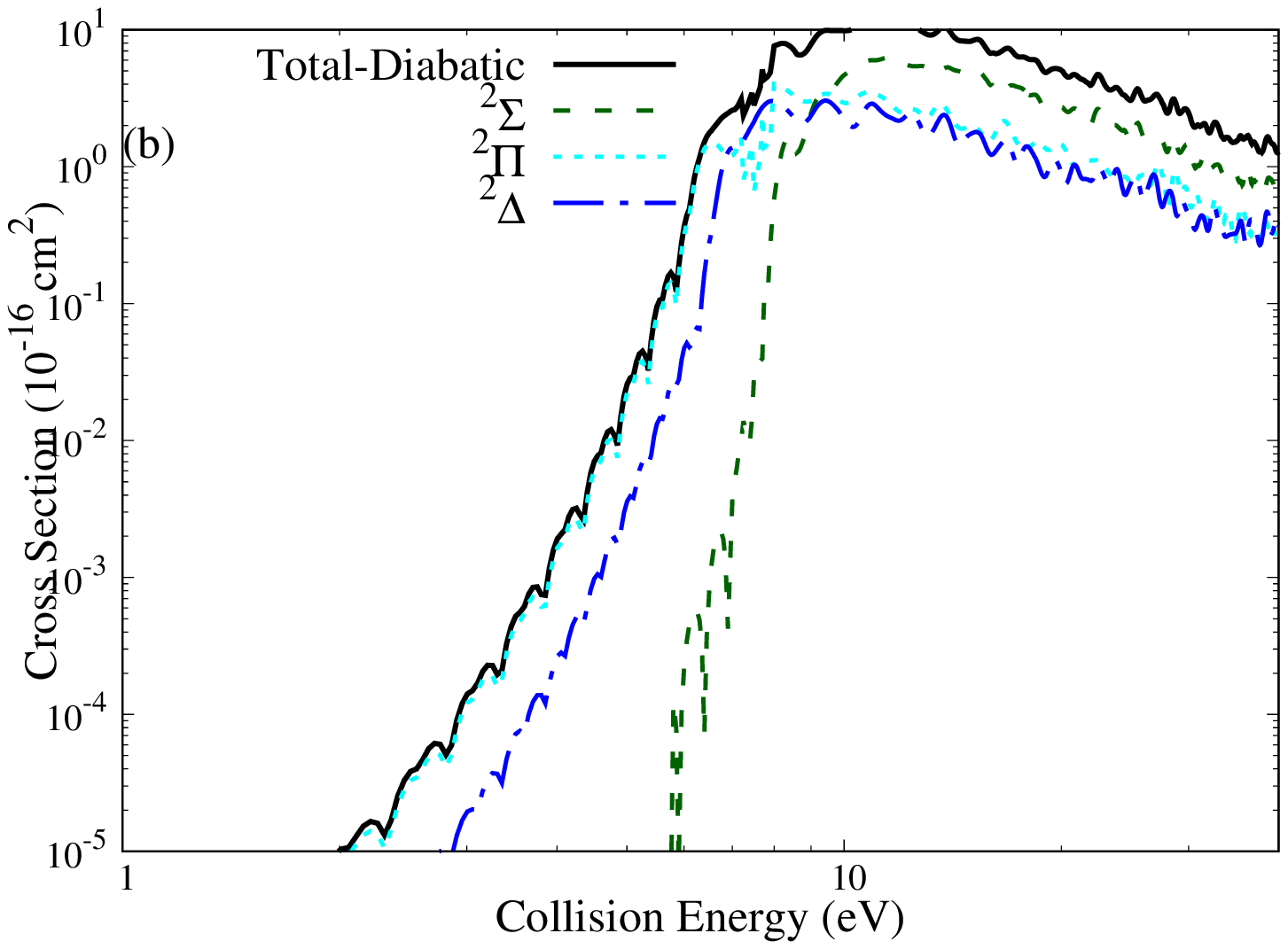}
\caption{\label{cross2} Calculated DR reaction cross section for $\mathrm{^4HeH^+}$ for the different symmetries, in the adiabatic (a) and diabatic (b) representations}
\end{figure}

Figure~\ref{cross2} shows the contributions of the individual electronic symmetries to the total DR cross section.

The results demonstrate that the recombination dynamics are dominated by states of $^2\Sigma$ symmetry over most of the investigated energy range. Above approximately 10 eV, the $^2\Sigma$ contribution accounts for the majority of the total cross section in both representations. This dominance can be attributed to the stronger electronic coupling between the $^2\Sigma$ resonant states and the ionic ground state, together with more favourable Franck–Condon overlap with the vibrational ground state of $\mathrm{HeH^+}$. Similar behaviour has been predicted previously for low-lying resonant states in $\mathrm{HeH^+}$ by Guberman~\cite{Guberman} and Larson \textit{et al.}~\cite{Larson15}.

At lower collision energies, however, the $^2\Pi$ states become increasingly important, particularly in the adiabatic representation. This enhanced contribution arises from strong rotationally induced mixing between the $^2\Sigma$ and $^2\Pi$ manifolds. The inclusion of rotational couplings therefore opens additional dissociative pathways that were absent in earlier calculations.

The contribution of the $^2\Delta$ symmetry remains relatively small throughout the investigated energy range. This behaviour is expected because $\Delta$ states are only indirectly connected to the ionic ground state through rotational interactions and therefore exhibit weaker effective coupling strengths.

The stronger dominance of the $^2\Sigma$ symmetry in the diabatic representation indicates that the diabatization procedure concentrates population transfer into a smaller number of strongly coupled channels. This observation confirms the important role played by the electronic coupling topology in determining dissociation probabilities.
\subsection{ Isotopic Effects in Dissociative Recombination}

\begin{figure}[!h]
\includegraphics[scale=0.35,angle=-90]{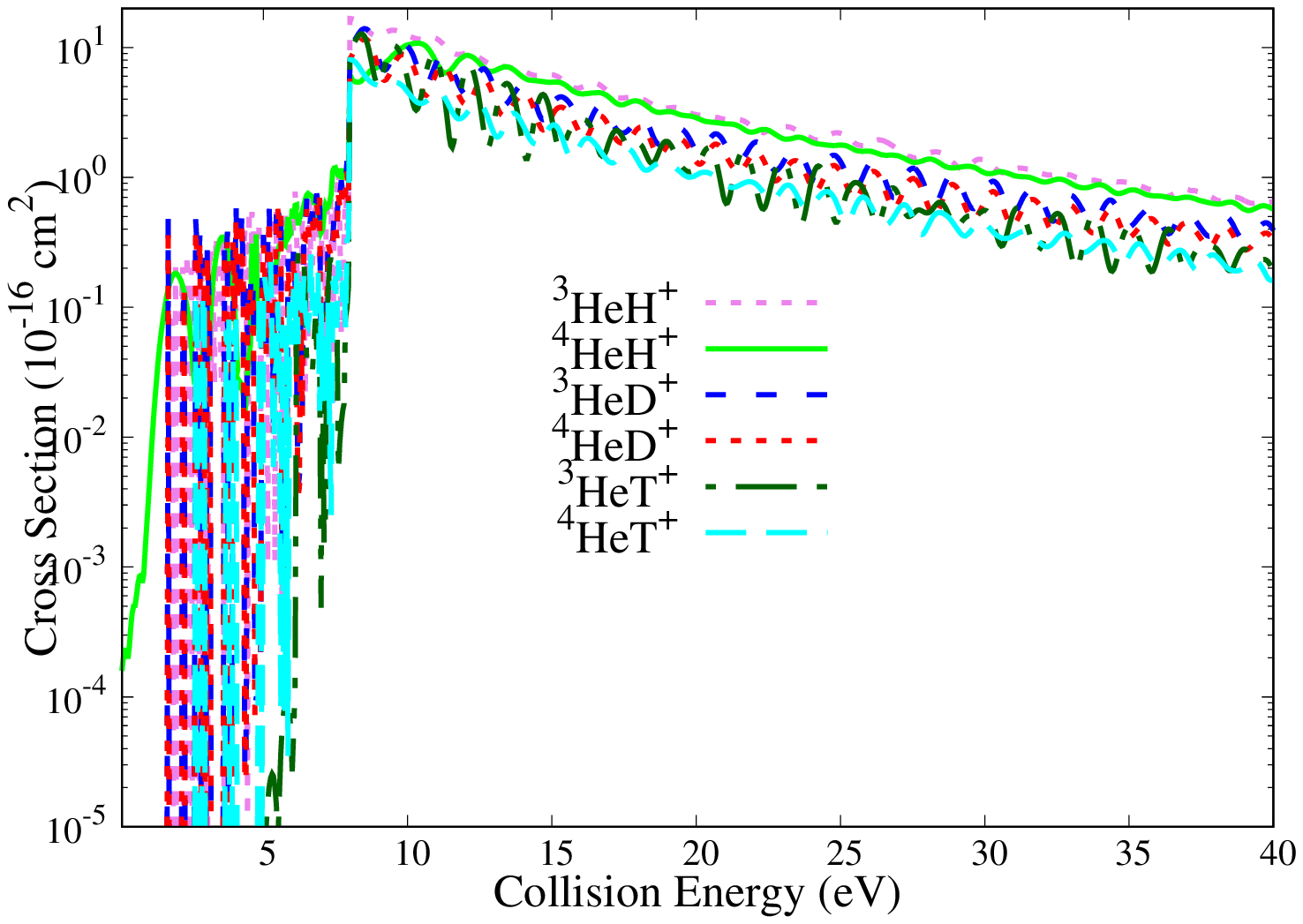}
\caption{\label{isot1} Total DR reaction total cross section for different $\mathrm{HeH^+}$ isotopologues. A clear dependence on reduced mass, particularly for collision energies above 5 eV.}
\end{figure}

The calculated DR cross sections for all isotopologues are shown in Figure~\ref{isot1}. A clear isotopic dependence is observed, with lighter isotopologues exhibiting systematically larger cross sections than heavier isotopologues.
The ordering
\begin{equation}
\mathrm{^3HeH^+} > \mathrm{^4HeH^+} > \mathrm{^3HeD^+}> \mathrm{^4HeD^+}> \mathrm{^3HeT^+}> \mathrm{^4HeT^+}\nonumber
\end{equation}

closely follows the inverse ordering of reduced masses.

This trend can be understood from the nuclear dynamics. For lighter isotopologues, the nuclei move more rapidly on the resonant potential-energy surfaces, increasing the probability that the system reaches the dissociation region before autoionization occurs. Consequently, the branching ratio favouring dissociation increases. Heavier isotopologues spend more time in the resonance region, allowing greater autoionization losses and reducing the overall DR probability.

Similar inverse mass scaling has been observed experimentally and theoretically in several molecular-ion systems, including isotopologues of $\mathrm{H_{2}^{+}}$ and $\mathrm{HeH^+}$~\cite{Larson98,Novotny2019}. The present calculations confirm that isotope effects remain significant even when a large manifold of coupled resonant states is included.

The separation between isotopologue cross sections becomes increasingly pronounced above approximately 5 eV. This suggests that the role of nuclear motion becomes more important at higher collision energies where a larger number of dissociative channels are energetically accessible.

\subsection{Resonant Ion-Pair Formation}
\begin{figure}[!h]
\includegraphics[scale=0.35,angle=-90]{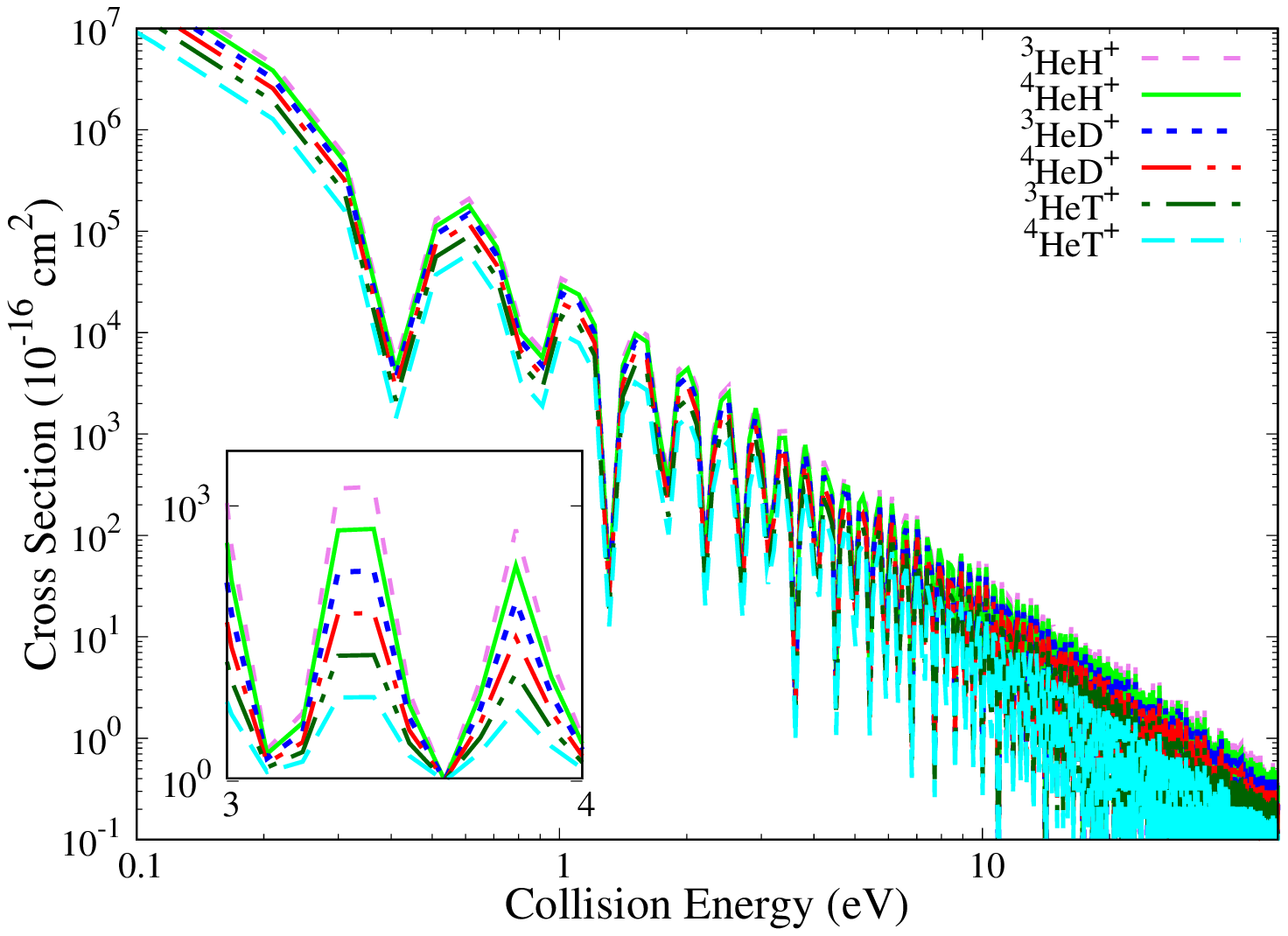}
\caption{\label{rip1} RIP formation total cross section for for different isotopes $\mathrm{He}$ and $\mathrm{H}$ when using \textbf{method 1}}
\end{figure}

\begin{figure}[!h]
\includegraphics[scale=0.35,angle=-90]{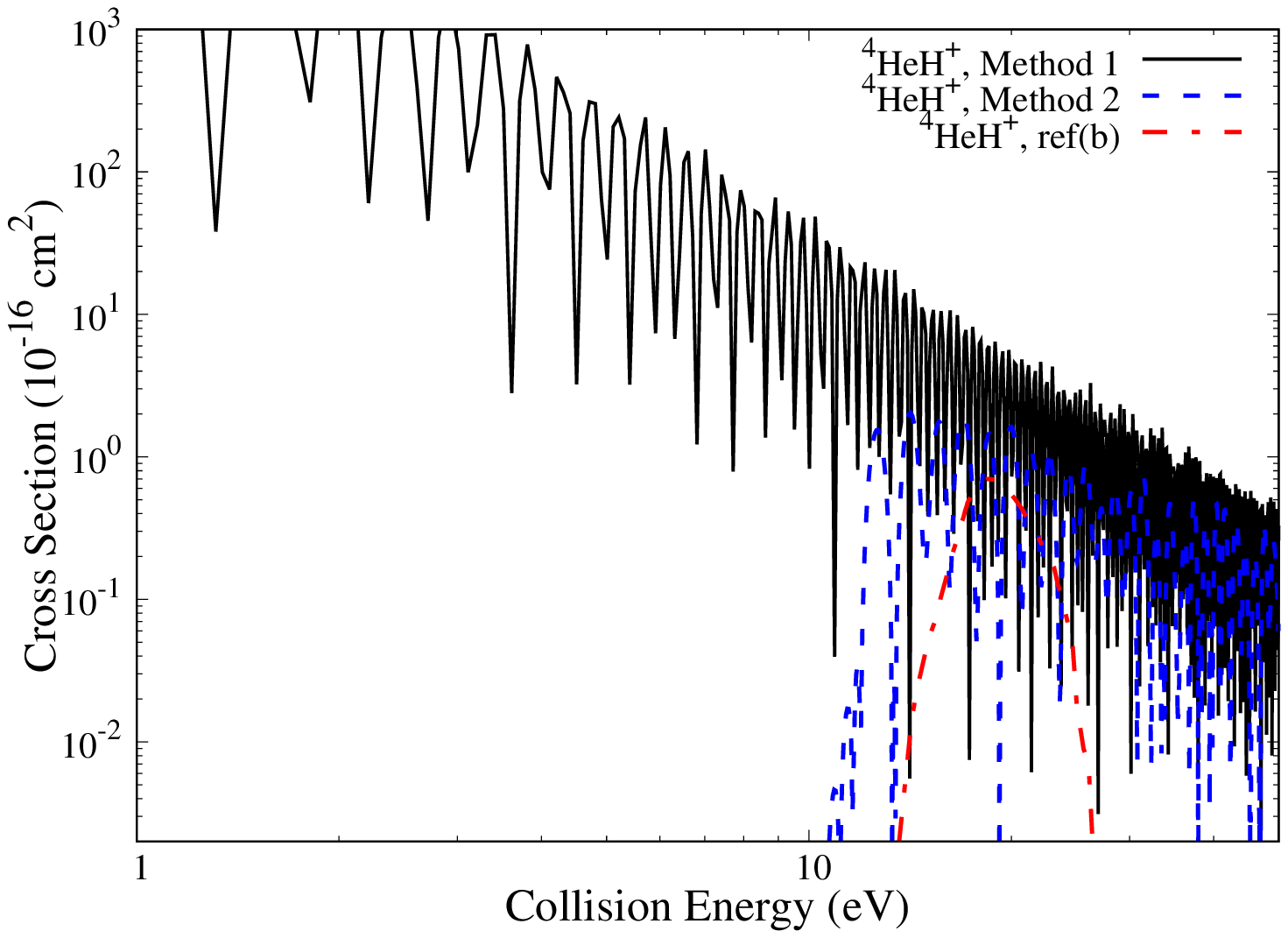}
\caption{\label{rip2} RIP formation total cross section for for different $\mathrm{^4He}$ and $\mathrm{H}$ when using \textbf{method 1} and \textbf{method 2} is compared with previous theoretical results from ref.(b)~\cite{Larson98}.}
\end{figure}

Figure~\ref{rip1} shows the resonant ion-pair (RIP) formation cross sections obtained using Method 1, while Figure~\ref{rip2} compares the two diabatization procedures.

A major result of the present work is that both diabatization approaches predict RIP cross sections significantly larger than those reported previously by Larson \textit{et al.}~\cite{Larson98}. The enhancement is particularly pronounced at low collision energies where previous calculations predicted almost negligible ion-pair formation.

The increased ion-pair yield can be attributed to three factors:

\begin{enumerate}
\item Inclusion of a substantially larger set of resonant states.
\item Explicit treatment of rotational couplings between different symmetries.
\item More complete treatment of avoided crossings and nonadiabatic transitions.
\end{enumerate}

Ion-pair formation is highly sensitive to the electronic coupling network because the process requires efficient population transfer from neutral covalent states into the ion-pair state. The strict 23-state diabatization employed in Method 2 preserves coupling across multiple avoided crossings simultaneously, thereby enhancing transfer probabilities into the ion-pair channel.

The comparison shown in Figure~\ref{rip2} demonstrates that Method 2 consistently predicts larger cross sections than Method 1 over nearly the entire energy range. This result indicates that two-state diabatization underestimates the cumulative effect of multiple avoided crossings. Similar conclusions have been reached in previous studies of ion-pair formation in molecular systems where multistate interactions play a dominant role ~\cite{Hedberg14,Larson16,Nkambule22}.

The RIP cross sections decrease rapidly at higher collision energies. Physically, this reflects the shorter interaction time available for nonadiabatic population transfer at higher relative velocities. As collision energy increases, direct dissociation and other competing channels become increasingly favourable.

\subsection{Isotopic Dependence of Ion-Pair Formation}
As seen in Figure~\ref{rip1}, the isotopic dependence observed for RIP formation mirrors that found for DR. Lighter isotopologues exhibit significantly larger ion-pair cross sections than heavier isotopologues.

This behaviour again reflects the influence of reduced mass on nonadiabatic dynamics. Lower reduced masses increase nuclear velocities and enhance the probability of traversing avoided crossings diabatically, thereby favouring population transfer into the ion-pair channel. The observed mass dependence therefore provides additional evidence that nonadiabatic coupling mechanisms govern the reaction dynamics.

The persistence of isotope effects in both DR and RIP channels suggests that reduced mass is a key parameter controlling electron-driven fragmentation processes in $\mathrm{HeH^+}$  isotopologues.

\section{\label{trates}Thermal Rate Coefficients}

While energy-dependent cross sections provide detailed information regarding the collision dynamics, plasma modelling and astrochemical simulations generally require temperature-dependent reaction rate coefficients. These are obtained by averaging the calculated cross sections over a Maxwell–Boltzmann distribution of electron energies~\cite{Novotny2019}.

For an electron temperature ($T_e$), the thermal rate coefficient is given by

\begin{equation}
k(T_e)=
\left(
\frac{8}{\pi m_e}
\right)^{\frac{1}{2}}
\frac{1}{\left(k_B T_e\right)^{\frac{3}{2}}}
\int_0^\infty
E\sigma(E)
e^{\left(
-\frac{E}{k_B T_e}
\right)}dE ,
\end{equation}

where ($m_e$) is the electron mass, ($k_B$) is the Boltzmann constant. This expression follows directly from averaging the product of the collision velocity and cross section over a Maxwellian electron energy distribution and has been widely employed in plasma and astrophysical modelling studies~\cite{Tennyson2010,Schneider2017}.

The present rate coefficients were obtained by numerically integrating the calculated cross sections over electron temperatures ranging from ($10^2$) to ($2 \times 10^4$) K. This temperature range was selected because it spans the conditions relevant to diffuse interstellar clouds, photodissociation regions, planetary nebulae, laboratory plasmas and primordial gas chemistry~\cite{GalliPalla1998,Lepp02,Schneider2017}.

For temperatures below approximately 100 K, the rate coefficients become increasingly sensitive to unresolved threshold resonances and indirect recombination pathways that are not explicitly included within the local complex potential treatment. Consequently, the present calculations are expected to be most reliable for temperatures above approximately 100 K, where direct dissociative recombination dominates and the wave-packet approach has been shown to provide accurate results~\cite{Orel95,Guberman,Tennyson2010}.
\subsection{Dissociative Recombination Rate Coefficients}
\begin{figure}[!h]
\includegraphics[scale=0.35,angle=-90]{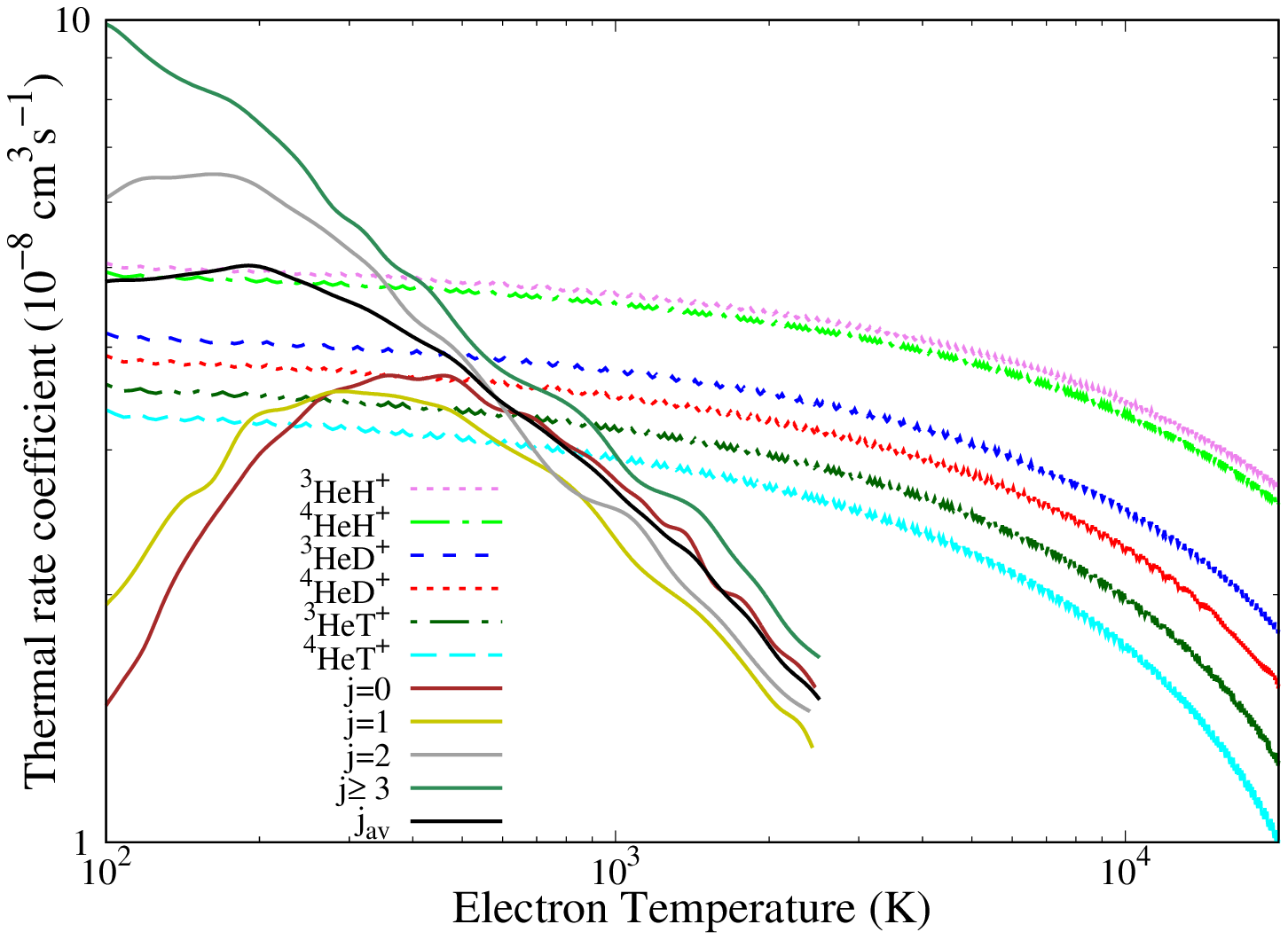}

\caption{Comparison of thermal rate coefficients for dissociative recombination (DR) \( \text{HeH}^+ \) for all isotopologues . The DR rate coefficient (dotted lines) is computed in the adiabatic representation including all 23 coupled states. The rate coefficients are compared with previous rotaional quantum state selective study (solid lines) from ref~\cite{Novotny2019}.
\label{rate1}}
\end{figure}

The thermal rate coefficients derived from the calculated DR cross sections are shown in Figure~\ref{rate1} for all $\mathrm{HeH^+}$ isotopologues. The figure also includes the rotational-state-dependent theoretical and experimental results reported by Novotny \textit{et al.}~\cite{Novotny2019}, corresponding to rotational levels $j=0$, $j=1$, $j=2$ and $j\ge3$.

For all isotopologues, the calculated rate coefficients decrease monotonically with increasing electron temperature. Such behaviour is characteristic of dissociative recombination processes and arises from the approximately inverse relationship between the recombination cross section and collision energy. As the temperature increases, the electron population shifts towards higher collision energies where the capture probability into dissociative resonant states becomes smaller, leading to reduced rate coefficients. Similar temperature dependences have been reported previously for molecular ions including $\mathrm{H_3^+}$, $\mathrm{HD^+}$, and $\mathrm{HeH^+}$ ~\cite{Guberman,Tennyson2010,Schneider2017}.

The present calculations reproduce the overall magnitude and temperature dependence observed in the rotationally resolved study of Novotný \textit{et al.}~\cite{Novotny2019}. In particular, the calculated rate coefficients lie within the envelope defined by the rotational populations $j=0$ to $j\ge3$ over most of the investigated temperature range. This agreement is significant because the present model does not explicitly resolve individual rotational states but instead includes rotational couplings between electronic symmetries through the nuclear dynamics Hamiltonian.

The comparison suggests that the dominant contribution to the thermal rate originates from the electronic dynamics rather than from detailed rotational-state effects. The remaining differences between the present results and those of Novotný \textit{et al.} can be attributed to the different theoretical treatments. The present work includes 23 coupled resonant states together with rotational couplings between $^{2}\Sigma$, $^{2}\Pi$, and $^{2}\Delta$ symmetries, whereas the rotationally resolved calculations were focused primarily on rotational excitation effects. Consequently, the present calculations generally predict somewhat larger rate coefficients because additional dissociative pathways become available through the extended manifold of resonant states.
This is what is observed at temperatures above 10$^3$ K.

Isotopic effect is also observed with lighter isotopes exhibiting a larger rate. This trend is consistent with the behaviour observed in the cross sections presented in Figure~\ref{isot1}. Lighter isotopologues possess larger nuclear velocities for a given collision energy, which increases the probability that the system dissociates before autoionization can occur. Similar isotope effects have been reported previously in both theoretical and experimental investigations of dissociative recombination~\cite{Larson98,Novotny2019}.

The separation between isotopologues becomes increasingly pronounced at higher temperatures. This behaviour reflects the stronger influence of nuclear dynamics at higher collision energies, where a larger number of resonant channels contribute to the recombination process. The persistence of the mass dependence over the entire temperature range indicates that nonadiabatic nuclear motion remains a controlling factor in the DR dynamics even when thermal averaging is performed.

\section{\label{concl}Conclusion}
In this work, we have carried out a comprehensive theoretical investigation of DR and RIP formation in $\mathrm{HeH^+}$ isotopologues using a time-dependent wave-packet approach. The nuclear dynamics were treated on a manifold of 23 coupled resonant electronic states of $^{2}\Sigma$, $^{2}\Pi$, and $^{2}\Delta$ symmetries, including rotational couplings between states of different electronic symmetries. Calculations were performed in both adiabatic and strictly diabatic representations, allowing a detailed assessment of the role of nonadiabatic interactions in the electron-collision dynamics.

The calculated DR cross sections are systematically larger than those reported in earlier theoretical studies by Sarpal \textit{et al.}~\cite{Sarpal94}, Orel \textit{et al.}~\cite{Orel95}, and Larson \textit{et al.}~\cite{Larson98}. The origin of this discrepancy can be attributed primarily to three factors. First, the present model includes a substantially larger manifold of resonant electronic states than previous calculations. Earlier wave-packet studies were generally restricted to a small subset of resonant states, whereas the present treatment includes 23 coupled states, thereby increasing the number of accessible dissociative pathways. Second, rotational couplings between $^{2}\Sigma$, $^{2}\Pi$, and $^{2}\Delta$ symmetries are included explicitly. These couplings provide additional channels for population transfer between resonant states and enhance the probability of dissociation before autoionization occurs. Third, the strict adiabatic-to-diabatic transformation developed by Larson \textit{et al.}~\cite{Larson16} preserves coupling across multiple avoided crossings simultaneously, allowing a more complete description of the electronic-state mixing that governs the dissociation dynamics.
 
The present calculations confirm that $^{2}\Sigma$ states dominate the recombination dynamics over most of the investigated collision-energy range. This behaviour is consistent with the strong coupling of these states to the ionic ground state and their favourable Franck–Condon overlap with the vibrational ground state of $\mathrm{HeH^+}$, as previously suggested by Guberman~\cite{Guberman} and Orel \textit{et al.}~\cite{Orel95}. In contrast, the $^{2}\Pi$ states contribute significantly at lower collision energies where rotationally induced mixing becomes important, while the contribution of $^{2}\Delta$ states remains comparatively small. The stronger dominance of $^{2}\Sigma$ channels observed in the diabatic representation demonstrates the important role played by electronic coupling topology in determining dissociation probabilities.

The calculated DR thermal rate coefficients exhibit the characteristic decrease with increasing electron temperature that has been observed experimentally and theoretically for molecular-ion recombination processes~\cite{Guberman,Sarpal1993,Schneider2017}. The overall temperature dependence is consistent with the rotational-state-resolved analysis reported by Novotný \textit{et al.}~\cite{Novotny2019}. In particular, the present rate coefficients fall within the envelope defined by the rotational populations $j = 0$, $j = 1$, $j = 2$, and $j \ge 3$, indicating that the present treatment captures the dominant physics governing electron capture and dissociation. The remaining differences in magnitude are likely due to the more extensive electronic-state manifold included in the present model and the explicit treatment of rotational couplings, both of which increase the total dissociation probability.

A clear isotopic dependence is observed for both DR and RIP processes. The reaction probabilities and thermal rate coefficients increase systematically with decreasing reduced mass.

This behaviour is consistent with the isotope trends reported by Larson \textit{et al.}~\cite{Larson98} and Novotný \textit{et al.}~\cite{Novotny2019} and can be understood in terms of faster nuclear motion in lighter isotopologues. The shorter residence time of the nuclear wave packet within the autoionization region increases the probability of dissociation relative to electron loss, thereby enhancing both the cross sections and the thermal rate coefficients. The persistence of this mass dependence after thermal averaging demonstrates that isotope effects are an intrinsic feature of the underlying nonadiabatic dynamics.

For resonant ion-pair formation, both diabatization schemes considered in the present work predict cross sections that are significantly larger than those obtained previously by Larson \textit{et al.}~\cite{Larson98}. The enhancement is particularly pronounced at low collision energies, where earlier calculations predicted negligible ion-pair production. The present results indicate that ion-pair formation is highly sensitive to the treatment of avoided crossings and electronic couplings. The strict 23 state diabatization yields larger cross sections than the two-state transformation because it preserves the cumulative effects of multiple avoided crossings and allows more efficient population transfer into the ion-pair channel. Similar sensitivity of ion-pair formation to electronic-state coupling has been reported in previous studies of $\mathrm{HeH}$ and related molecular systems~\cite{Hedberg14,Larson16,Nkambule22}.

The astrophysical implications of these findings are significant. Since $\mathrm{HeH^+}$ is widely regarded as the first molecular ion formed in the Universe~\cite{GalliPalla1998,Lepp02,Black76,Stromholm,Guesten19}, accurate DR and RIP data are essential for modelling primordial chemistry, diffuse interstellar clouds, photodissociation regions, planetary nebulae, and other ionized astrophysical environments. The larger DR rate coefficients predicted here suggest more efficient destruction of $\mathrm{HeH^+}$ than assumed in some earlier astrochemical models based on the rates of Galli and Palla ~\cite{GalliPalla1998} and Larson \textit{et al.}~\cite{Larson98}. Consequently, updated abundance calculations may be required to assess the impact of these revised reaction rates on the chemical evolution of primordial gas and the interpretation of astronomical observations such as the detection of $\mathrm{HeH^+}$ by Gusten \textit{et al.}~\cite{Guesten19}.

The present calculations employ the local complex potential (boomerang) approximation for the treatment of autoionization~\cite{BirtwistleHerzenberg1971,Herzenberg68,Herzenberg79} . Although this approach has been successfully applied in numerous studies of dissociative recombination~\cite{Orel95,Hedberg14,Guberman}, it neglects interference effects between overlapping resonances and does not explicitly include indirect recombination through high-lying Rydberg states. More rigorous approaches based on multichannel quantum defect theory (MQDT)~\cite{Guesten19,GiustiSuzor1980} or modern R-matrix methodologies~\cite{Sarpal1993,Tennyson2010,Schneider2017} may therefore provide useful benchmarks for assessing the remaining theoretical uncertainties. Nevertheless, the overall agreement with available experimental measurements~\cite{Mowat95,Sundstrom94,Yousif89,Habs89} and the rotationally resolved results of Novotný \textit{et al.}~\cite{Novotny2019} suggests that the present model captures the dominant mechanisms governing electron-driven fragmentation of $\mathrm{HeH^+}$.


\section*{Acknowledgemets}
Electronic structure  and electron scattering data used in this study was generously provided by Prof. \AA sa Larson from Stocholm University and Prof. Ann E. Orel from  the University of Carlifonia, Davis.
\section*{Declarations}

\subsection*{Data Availability Statement}
Data used in this work can be made available at request. Data Generously provided by Prof. \AA. Larson and Prof. A. E. Orel was analysed using our own FORTRAN codes. Derived data supporting the findings of this study are available from the corresponding author upon reasonable request.
\subsection*{Author Contribution}
All authors contributed to the study conception and design. Material preparation, data collection and analysis were performed by Sifiso M. Nkambule, Malibongwe Tsabedze, Oscar N. Mabuza and Mbuso K. Matfunjwa. The first draft of the manuscript was written by Sifiso M. Nkambule and all authors commented on previous versions of the manuscript. All authors read and approved the final manuscript.
\subsection*{Funding}
No funding was received for conducting this study.
\subsection*{Financial Interests}
The authors have no conflicts of interest to declare that are relevant to the content of this article.
\normalem
\pagebreak
\bibliography{references.bib}
\end{document}